\documentclass[12pt]{article}
\usepackage[pdftex]{graphicx}
\usepackage{geometry,hyperref,xcolor}
\hypersetup{
    colorlinks=true,
    linkcolor={red!50!black},
    citecolor={blue!50!black},
    urlcolor={blue!80!black},
    pdfinfo={
      Title={First Run of the LArIAT Testbeam Experiment},
      Author={William Foreman},
      Subject={LArIAT},
      Keywords={}
  	}
}

\newgeometry{left=1.2in,right=1.2in,top=1.5in,bottom=1.5in}

\newcommand\pubnumber{DPF2015-302}
\newcommand\pubdate{\today}

\def\uchi{Enrico Fermi Institute\\
University of Chicago, 5640 S Ellis Ave, Chicago, IL 60637}
\def\support{\footnote{Work supported by the National Science Foundation, U.S. DOE Office of Science, Fermi National Accelerator Laboratory, and the Enrico Fermi Institute at the University of Chicago.}}

\def\Title#1{\begin{center} {\Large #1 } \end{center}}
\def\Author#1{\begin{center}{ \sc #1} \end{center}}
\def\Address#1{\begin{center}{ \it #1} \end{center}}

\newcommand\pubblock{\rightline{\begin{tabular}{l} \pubnumber\\
         \pubdate  \end{tabular}}}
\newenvironment{Abstract}{\begin{quotation}  }{\end{quotation}}
\newenvironment{Presented}{\begin{quotation} \begin{center} 
             PRESENTED AT\end{center}\bigskip 
      \begin{center}\begin{large}}{\end{large}\end{center} \end{quotation}}

\def\beq{\begin{equation}}
\def\eeq#1{\label{#1}\end{equation}}
\def\eeqn{\end{equation}}

\def\beqa{\begin{eqnarray}}
\def\eeqa#1{\label{#1}\end{eqnarray}}
\def\eeqan{\end{eqnarray}}

\let\bar=\overbar

\def\etal{{\it et al.}}

\def\Dslash{\not{\hbox{\kern-4pt $D$}}}
\def\dslash{\not{\hbox{\kern-2pt $\del$}}}

\def\msb{{\bar{\ssstyle M \kern -1pt S}}}

\begin{document}
\begin{titlepage}
\pubblock

\vfill
\Title{First Run of the LArIAT Testbeam Experiment}
\vfill
\Author{William Foreman\support \hspace{0.0ex} for the LArIAT Collaboration}
\Address{\uchi}
\vfill
\begin{Abstract}
LArIAT (Liquid Argon In A Testbeam) aims to characterize the response of a liquid argon time projection chamber (LArTPC) to the particles often seen as final-state products of $\sim$1 GeV neutrino interactions in existing and planned detectors.  The experiment uses the ArgoNeuT cryostat and its refurbished 170-liter-active-volume TPC placed in a tunable tertiary beamline produced from a high-energy pion beam at the Fermilab Test Beam Facility (FTBF).  The TPC was modified to accommodate cold readout electronics and a light collection system. The first run took place May-June of 2015, and the collected data will help in understanding electron recombination behavior, shower reconstruction, particle identification, muon sign determination, pion and kaon interactions in argon, and the use of scintillation light for calorimetry.
\end{Abstract}
\vfill
\begin{Presented}
DPF 2015\\
The Meeting of the American Physical Society\\
Division of Particles and Fields\\
Ann Arbor, Michigan, August 4--8, 2015\\
\end{Presented}
\vfill
\end{titlepage}
\def\thefootnote{\fnsymbol{footnote}}
\setcounter{footnote}{0}

\section{Introduction}

Liquid argon time projection chambers (LArTPCs) use planes of angled wires to detect ionization electrons drifted along a uniform electric field.  Their scalability, low energy threshold, calorimetric resolution, and ability to reconstruct particle interactions in 3D bubble-chamber-like quality make them ideal for the detection of the charged products from neutrino interactions  -- primarily muons, electrons, pions, and protons.

Following decades of technical development in Europe, LArTPC technology was first demonstrated on a large scale in 2010 by ICARUS, which successfully operated a deep-underground detector filled with 760 tons of liquid argon at INFN Gran Sasso Laboratory in Italy~\cite{icarus}.  In parallel, the ArgoNeuT Collaboration collected thousands of neutrino and antineutrino events using a 0.75-ton detector placed along the NuMI beamline at Fermilab from 2009-2010~\cite{argoneut}.  MicroBooNE, a 170-ton LArTPC along the Booster neutrino beamline at Fermilab, is now beginning its first physics run and will be joined by two additional LAr detectors in the coming years as part of a comprehensive short-baseline neutrino oscillation program~\cite{microboone,sbn}.  The massive DUNE detector will also use LArTPC technology to observe neutrinos fired through the Earth from Fermilab toward a deep underground site in South Dakota~\cite{dune}.

With interest in liquid argon surging among neutrino physicists, a dedicated calibration of this emerging technology is invaluable.  The LArIAT (LArTPC In A Testbeam) experiment fills this niche by operating the refurbished ArgoNeuT detector in a tunable test beam at Fermilab.  LArIAT explores particle response calibration, technical R\&D, and several physics topics relevant to current and future liquid argon detectors~\cite{lariat}.

\section{Tertiary Beam}

Protons accelerated to 120 GeV in Fermilab's Main Injector are used to create secondary beams of charged particles directed toward the Fermilab Test Beam Facility (FTBF)~\cite{ftbf}.  This secondary beam, composed mostly of high-energy pions and protons from 8 to 32 GeV, is impinged onto a copper target embedded in a steel 12$^o$ collimator to produce an outgoing ``tertiary'' beam housed in FTBF's MCenter enclosure.  

\begin{figure}[htb]
\centering
\includegraphics[width=0.95\textwidth]{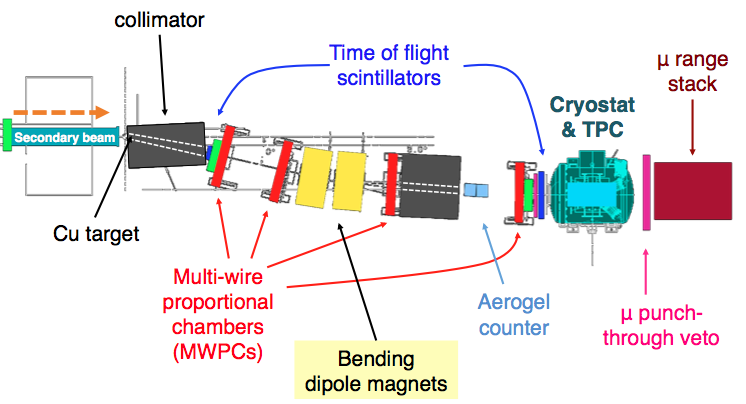}
\caption{A schematic diagram of LArIAT's tertiary beamline in the MCenter enclosure at Fermilab.}
\label{fig:beamline}
\end{figure}

As shown in Figure~\ref{fig:beamline}, bending magnets and four wire chambers positioned along the tertiary beamline allow the measurement of particles' momenta while time-of-flight (TOF) scintillators separate pions and muons from the heavier (and slower) protons and kaons.  Aerogel Cherenkov detectors and a muon range stack help to further distinguish pions from muons.  

The beam's momenta spectra and composition is tuned by adjusting the field intensity in the bending magnets (see Table~\ref{tab:composition}).  The momentum profile ranges from 0.2 to 2.0 GeV, overlapping the range spanned by particles emerging from neutrino events in the NuMI and Booster beamlines as well as those expected in DUNE. \cite{lariat}

The tertiary beamline detectors were installed and commissioned in the summer of 2014, and were then used to characterize the beam in preparation for physics running.

\begin{table}[tb!]
\begin{center}
\begin{tabular}{c|cc}  
Particle type  &  +0.14 Tesla field & +0.35 Tesla field \\ \hline
  $\pi+$       &   32.6\%           &           59.2\%  \\
  $e+$         &   55.2\%           &           13.3\%  \\
  $\gamma$     &    4.0\%           &           2.3\%     \\
  $p+$         &    6.7\%           &           23.5\%     \\
  $\mu+$       &   1.9\%            &           1.9\%  \\
  $K+$         &    0.005\%         &           0.06\%  \\
\end{tabular}
\caption{Particle species reaching the TPC for two different field settings in the tertiary beamline's bending magnets for an 8 GeV secondary beam, as predicted by Monte Carlo simulations~\cite{lariat}.}
\label{tab:composition}
\end{center}
\end{table}

\section{TPC and Light Collection System}

ArgoNeuT's TPC, with an active volume of 170 liters (47 {\it w} $\times$ 40 {\it h} $\times$ 90 {\it l} cm$^3$), was refurbished for use in LArIAT to accommodate cold readout electronics.  Three new wire planes were installed as well, though their geometry remains unchanged. Drifting ionization electrons first pass a non-instrumented plane of vertical wires that helps shape the electric field near the wires and shields them from induction due to drifting charge.  The next two planes feature wires angled at +/-60$^o$ relative to the beam direction and record induction and charge collection signals.  The wire spacing is 4mm for each plane.

A side access flange on the cryostat was instrumented to allow for a light detection system.  Two cryogenic, high-quantum efficiency PMTs as well as three silicon photomultipliers (SiPMs) on custom preamplifier boards are supported by a PEEK structure mounted to the flange.  The photosensitive windows of these devices are held 2-3cm behind the wireplanes and peer into the active volume of the TPC.

\begin{figure}[tb!]
\centering
\includegraphics[height=2.65in]{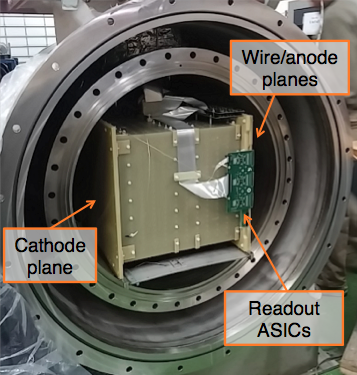}
\hspace{0.5cm}
\includegraphics[height=2.65in]{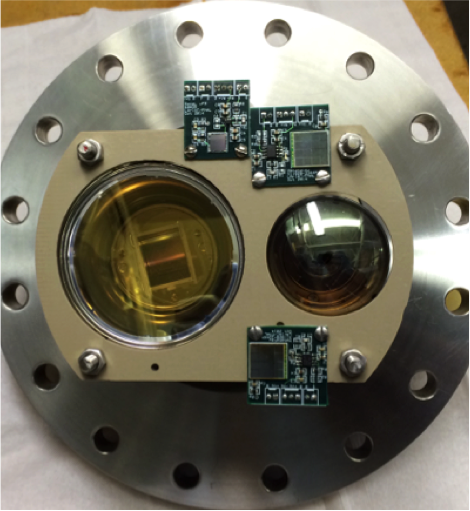}
\caption{The LArIAT TPC inside the ArgoNeuT vacuum-insulated cryostat (left) and the light collection system mounted to the side flange (right).}
\label{tpc}
\end{figure}

Reflective dielectric substrate foils coated in a thin layer of tetraphenyl butadiene (TPB) line the TPC's four field cage walls.  Vacuum-ultraviolet photons from LAr scintillation interact with the TPB to induce re-emission of visible light that is reflected back into the active volume to be detected.  This technique increases light collection efficiency and spatial uniformity relative to traditional light collection systems in LArTPCs where the wavelength-shifting occurs at transparent TPB-coated disks suspended in front of each PMT.

\section{Summary of Run I}

The cryostat was filled completely by April 29, 2015, and data-taking took place April 30 to July 7.  A total of 9 weeks of beam data was collected followed by a week of experimental runs using several low-energy radioactive sources.  Each $\sim$1-minute accelerator supercycle includes 4 seconds of delivered beam.  This is followed by a 24-second window during which triggered cosmic ray events are recorded using dedicated scintillation paddles and a light-based trigger to tag stopping and decaying muons.

\begin{figure}[tb]
\centering
\includegraphics[width=0.65\textwidth]{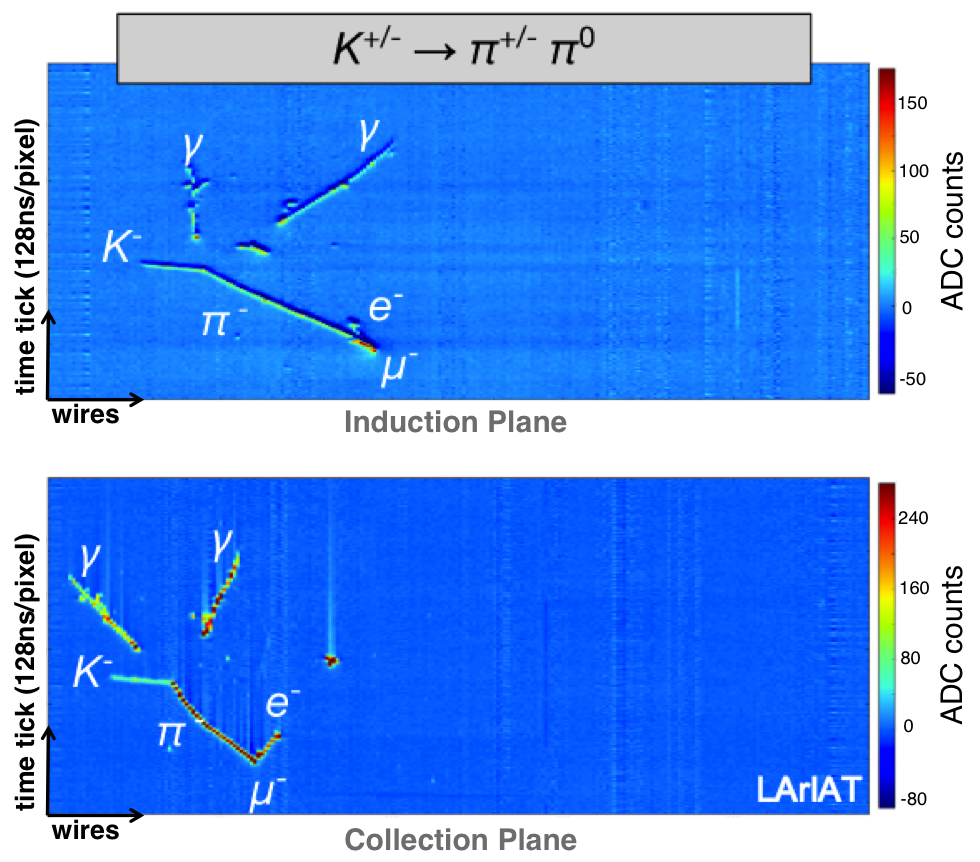}
\caption{An example of a real LArIAT data event showing a candidate kaon decay.  The beam travels parallel to $\hat{z}$, entering the TPC at the upstream face (left edge of display).  Ionization charge drifts toward the wires along $-\hat{x}$ (downward-vertical direction of display).  Full drift period is approximately 384$\mu$s.}
\label{event}
\end{figure}

The first 3 weeks consisted mostly of beam tuning and optimization of the data acquisition.  Following this period, $\sim$44k spills were collected under optimal beam conditions, with $\sim$5-10 analyzable beam events of varying particle types per each 4-second spill.

Data was collected using different combinations of three secondary beam energies (16, 32, 64 GeV) and four tertiary beam magnet current settings within $\pm$20-100A to ensure the widest possible range of particle species and momenta were recorded.

\section{Planned and Ongoing Analyses}

LArIAT offers analyses covering detector calibration as well as physics topics relevant to current and upcoming LArTPC-based precision neutrino experiments.  

Each analysis, to varying degrees, relies on software-based reconstruction of signals from the beamline detectors, TPC, and light collection system.  A custom version of LArSoft -- a detector-generic suite of algorithms tailored for LArTPCs~\cite{larsoft} -- has been developed for LArIAT.  Features are regularly being added and improved as progress is made toward fully-automated data reconstruction.

Foundational calibration studies include measuring the drift electron lifetime, electronics response calibration, and charge recombination as a function of electric field.  The primary higher-level physics analyses are outlined below:

\begin{itemize}

\item {\bf $e/\gamma$ separation.}  In neutrino experiments, charged-current (CC) $\nu_e$ interactions produce an outgoing $e^{+/-}$ while neutral-current (NC) interactions often produce a $\pi^0$ decaying promptly to two $\gamma$'s, each of which can pair-produce to form electromagnetic showers.  Distinguishing electron showers from $\gamma$ showers is possible in LArTPCs, though an experimental measure of the efficiency of this technique has never been carried out.

\item {\bf $\pi$-Ar interactions.}  The lack of knowledge on $\pi$-Ar cross sections is a major source of uncertainty in neutrino oscillation experiments.  The $\pi$-rich data from LArIAT will help in developing $\pi$ identification algorithms based on their interaction modes in argon.  Measuring the inclusive $\pi$-Ar cross section as well as some of its contributing components -- elastic and inelastic scattering, absorption, charge exchange, and secondary pion production -- is a highly active area of analysis.  Pion single charge exchange reactions (for example, $\pi^+ + n \rightarrow \pi^0 + p$) are particularly important for future neutrino oscillation searches as the neutral pion decay ($\pi^0 \rightarrow \gamma\gamma$) can mimic the signature of a NC-$\nu$ event.
    
\item {\bf Non-magnetic $\mu$ sign determination.}  In LAr, stopping $\mu^+$ (a product of CC-$\bar{\nu_\mu}$ events) always decay to $e^+$, while $\mu^-$ (from CC-$\nu_\mu$ events) are captured by Ar nuclei $\sim$76\% of the time without getting a chance to decay.  The detection of a delayed Michel electron (or the $\gamma$ and neutron byproducts of a nuclear capture) using a combination of topology and scintillation light information~\cite{sorel} is a useful handle in determining the charge of final-state muons in neutrino events -- potentially allowing for the selection of statistically-pure $\nu$ or $\bar{\nu}$ event samples.  Ongoing analysis and reconstruction of Michel electrons in LArIAT will serve as a precursor to systematic $\mu$ sign determination studies.

\item {\bf Kaon interactions.}  Future proton decay searches in massive LArTPCs like DUNE will require the efficient detection of kaons.  A small fraction of events in LArIAT are expected to be kaons, enabling the possibility of the first systematic study of their interactions in a LArTPC and how to identify them (see Figure~\ref{event}).

\item {\bf Scintillation light.} A wealth of information about a particle's deposited energy and ionization density is encoded in LAr scintillation.  Analyses are underway to enhance calorimetry and particle identification capabilities using this light.
 
\end{itemize}

\section{Conclusion}

The LArTPC has become the choice detector for the foreseeable future of precision neutrino physics, and LArIAT offers an unprecedented chance to characterize the performance of this technology in a controlled environment.  The range of particles and energies produced in LArIAT's tertiary beamline are of prime relevance not only to MicroBooNE and the upcoming short-baseline program at Fermilab, but to long-baseline neutrino physics to be explored by DUNE.

Reconstruction and analysis is well underway on data collected during LArIAT's Run I.  Results are expected to bolster advancements in particle ID and event reconstruction in liquid argon while also contributing useful physics that will aid analyses in other LArTPCs.

Run II will begin in February of 2016 after some minor changes, including cryogenic phase separator improvements, bias voltage card modifications to eliminate cross-talk between wireplanes, repair of a faulty PMT base, and installation of new larger SiPMs mounted to custom readout boards.



\begin{thebibliography}{99}


\bibitem{icarus}
ICARUS Collaboration (C. Rubbia \etal), {\it Underground operation of the ICARUS T600 LAr-TPC: first results}, JINST 6 P07011, \href{http://arxiv.org/abs/1106.0975}{arXiv:1106.0975} [hep-ex] (2011).

\bibitem{argoneut}
ArgoNeuT Collaboration (C. Anderson \etal), {\it The ArgoNeuT Detector in the NuMI Low-Energy beam line at Fermilab}, JINST 7 P10019, \href{http://arxiv.org/abs/1205.6747}{arxiv:1205.6747} [physics.ins-det] (2012).

\bibitem{microboone}
MicroBooNE Collaboration, {\it The MicroBooNE Technical Design Report} (2012).

\bibitem{sbn}
R. Acciarri \etal, {\it A Proposal for a Three Detector Short-Baseline Neutrino Oscillation Program in the Fermilab Booster Neutrino Beam}, \href{http://arxiv.org/abs/1503.01520}{arXiv:1503.01520} [physics.ins-det] (2015).

\bibitem{dune}
LBNE/DUNE Collaboration (T. Akiriet \etal), {\it The 2010 Interim Report of the Long-Baseline Neutrino Experiment Collaboration Physics Working Groups}, \href{http://arxiv.org/abs/1110.6249}{arXiv:1110.6249} [hep-ex] (2011).

\bibitem{lariat}
F. Cavanna, M. Kordosky, J. Raaf, B. Rebel on behalf of the LArIAT Collaboration, {\it LArIAT: Liquid Argon In A Testbeam}, \href{http://arxiv.org/abs/1406.5560}{arXiv:1406.5560} [physics.ins-det] (2014).

\bibitem{ftbf}
FTBF (Fermilab Test Beam Facility), http://ftbf.fnal.gov.

\bibitem{larsoft}
LArSoft collaboration, E.  Church, {\it LArSoft: a software package  for  liquid  argon  time  proportional  drift  chambers}, \href{http://arxiv.org/abs/1311.6774}{arXiv:1311.6774} [physics.ins-det] (2013).

\bibitem{sorel}
M. Sorel, {\it Expected performance of an ideal liquid argon neutrino detector with enhanced sensitivity to scintillation light}, JINST 9 P10002, \href{http://arxiv.org/abs/1405.0848}{arXiv:1405.0848} [physics.ins-det] (2014).

\end{thebibliography}
\end{document}